\documentclass[12pt]{article}
\usepackage{amsmath}
\usepackage{epsfig}
\usepackage{amssymb,amsfonts}

\advance\voffset by -2.0 cm
\advance\hoffset by -1.25cm
\newcommand{\be}{\begin{equation}}
\newcommand{\ee}{\end{equation}}
\newcommand{\bea}{\begin{eqnarray}}
\newcommand{\eea}{\end{eqnarray}}

\newcommand{\D}{\Delta}

\newcommand{\N}{\mathcal{N}}

\newcommand{\bt}[1]{{\bar t}}

\newcommand{\thp}{\theta^{+}}
\newcommand{\thm}{\theta^{-}}
\newcommand{\thpi}[1]{\theta^+_{#1}}

\newcommand{\epp}{\epsilon_+}
\newcommand{\epm}{\epsilon_-}

\newcommand{\thNp}{\theta^{(+)}}
\newcommand{\thNm}{\theta^{(-)}}
\newcommand{\bthNp}{\bar{\theta}^{(+)}}
\newcommand{\bthNm}{\bar{\theta}^{(-)}}
\newcommand{\thNpm}{\theta^{(\pm)}}
\newcommand{\bthNpm}{\bar{\theta}^{(\pm)}}

\def\Q{{\cal Q}}

\def\D{{\cal D}}
\newcommand{\beqa}{\begin{eqnarray}}
\newcommand{\eeqa}{\end{eqnarray}}
\newcommand{\beq}{\begin{equation}}
\newcommand{\eeq}{\end{equation}}

\title{\vspace{2cm}\LARGE \textbf{Manifestly Supersymmetric RG Flows}
\vspace*{0.5cm}}

\author{
Matthias R.\ Gaberdiel\thanks{\tt E-mail: gaberdiel@itp.phys.ethz.ch}\,\, and  
Stefan Hohenegger\thanks{\tt E-mail: stefanh@itp.phys.ethz.ch}
\\ \\
{\small Institut f\"ur Theoretische Physik, 
ETH Z\"urich} \vspace*{-0.1cm} \\
{\small 8093 Z\"urich, Switzerland} \vspace{0.3cm} \\
}
\date{\today}
\begin{document}
\maketitle
\thispagestyle{empty}
\begin{abstract}
Renormalisation group (RG) equations in two-dimensional $\N=1$ supersymmetric 
field theories with boundary are studied. It is explained how a manifestly $\N=1$
supersymmetric scheme can be chosen, and within this scheme the RG equations 
are determined to next-to-leading order. We also use these results to revisit the question
of how brane obstructions and lines of marginal stability appear
from a world-sheet perspective. 
\end{abstract}

\newpage
\renewcommand{\theequation}{\arabic{section}.\arabic{equation}}




\section{Introduction}

The behaviour of supersymmetric D-branes under deformations of the closed
string background has a surprisingly rich and interesting structure. For example, 
even if the closed string remains supersymmetric under the deformation, the same 
may not be true in the presence of a D-brane. If this is the case one says that the deformation is 
`obstructed' by the D-brane. Another phenomenon that has attracted
a lot of attention recently \cite{Denef:2007vg,Gaiotto:2008cd,Gaiotto:2009hg} concerns lines 
of marginal stability: a supersymmetric brane may decay into a superposition of 
(supersymmetric) D-branes as the string background is modified (in a supersymmetric 
fashion). 

{}These phenomena were recently studied from the point of view of the world-sheet
\cite{Brunner:2009mn}. From this perspective obstructions are associated with
bulk perturbations for which the (adjusted) boundary condition does not preserve
the $\N=2$ super(conformal) symmetry any longer
(see also \cite{Fredenhagen:2006dn,Fredenhagen:2007rx}). Indeed, the supervariation 
of these perturbations vanishes only up to total (bosonic) derivatives, which in the presence 
of a D-brane generically lead to non-vanishing boundary contributions
\cite{Warner:1995ay,Hori:2004ja}. The latter can sometimes be cancelled by adding 
new boundary terms to the sigma model action, but this is not necessarily possible
and the D-brane may therefore be obstructed.
However, it is always possible to add a suitable boundary term so that an
$\N=1$ subalgebra is preserved. This is important from the point of view of string theory 
since the $\N=1$ algebra describes a gauge symmetry in this context (and hence must be 
preserved for consistency). Furthermore, lines of marginal stability appear when this boundary 
deformation (that is added to preserve the $\N=1$ supersymmetry) becomes 
relevant. 
\smallskip

Given that it is always possible to preserve an $\N=1$ supersymmetry
it should be possible to formulate the combined bulk and boundary deformation
problem in a {\em manifestly $\N=1$ supersymmetric} fashion. In fact, as we shall
explain in this paper, the boundary correction term has a natural interpretation in terms
of a superspace description of the problem. This observation can be used 
to formulate a renormalisation group scheme in which the $\N=1$
supersymmetry is (manifestly) preserved. Given what we said above, 
this is a very natural scheme for superstring calculations.

Within this scheme we then analyse the coupled renormalisation group equations,
thereby combining the superspace approach of {\it e.g.}~\cite{Brunner:2009mn} with 
methods of perturbed conformal field theory (see in
particular \cite{Fredenhagen:2006dn,Gaberdiel:2008fn}). Among other things 
we identify the precise coupling constant which controls the bulk induced boundary RG source 
term of \cite{Fredenhagen:2006dn}, and we explain how the change in conformal dimension 
of a boundary field can be calculated \cite{Gaberdiel:2008fn} in this context. We also apply 
these techniques to the case of (cc) perturbations of B-type branes. In our supersymmetric
scheme the first order bulk induced boundary RG source term always vanishes
for marginal boundary fields. However, the (cc) bulk perturbation may change
the conformal dimension of a marginal boundary field, and thus induce an instability. 
Finally, we compare these findings with results that had been obtained previously
using matrix factorisation techniques \cite{Baumgartl:2007an} (see also
\cite{Baumgartl:2008qp,Jockers:2008pe,Alim:2009rf,Alim:2009bx,Grimm:2009ef,Aganagic:2009jq} 
for related recent work). In particular,
we show (at least in an example) that the boundary field that becomes relevant
is precisely the one predicted from the analysis of  \cite{Baumgartl:2007an}. 
\smallskip

For the case of a (ca) perturbation of a B-type brane, on the other hand, one does
not expect any obstructions, and thus generically the full $\N=2$ supersymmetry 
should be preserved. This suggests that the (ca) deformation problem should 
have a manifestly $\N=2$ supersymmetric formulation, and this again turns out to be true. 
However, there {\em exist} lines of marginal stability in this context. They are associated
to a breakdown of this manifestly $\N=2$ supersymmetric scheme. 
\medskip

Supersymmetric D-branes have been studied extensively, using sigma model techniques, 
in the past, see for example 
\cite{Ooguri:1996ck,Hanany:1997vm,Hori:2000ck,Hori,Albertsson:2001dv,Albertsson:2002qc,%
Lindstrom:2002jb}. 
More recently, a manifestly $\N=1$ and $\N=2$ supersymmetric formulation for D-branes 
has also been given and interpreted in terms of generalised geometry
\cite{Koerber:2003ef,Sevrin:2007yn,Sevrin:2008tp,Sevrin:2009na}. 
Here we study how the manifestly $\N=1$ supersymmetric formulation can be 
maintained under supersymmetric bulk deformations.
\bigskip

The paper is organised as follows. In section~\ref{Sect:ModelSuperspace} we explain
how to formulate the deformation problem in a manifestly $\N=1$ supersymmetric
fashion. In section~\ref{Sect:RGequations}  we determine the RG equations 
in the associated scheme. These results are then applied to (cc) perturbations of 
B-type branes in section~\ref{Sec:cc}. In section~\ref{ca}, 
we explain how a manifestly $\N=2$ supersymmetric description is available for (ca) 
perturbations of B-type branes, and section~5 contains our conclusions.
 There are a number of appendices in which some of the
more technical material has been collected.


\section{Manifestly  supersymmetric theories with boundary}\label{Sect:ModelSuperspace}

In this section we want to discuss manifestly $\N=1$ supersymmetric field theories 
on two-dimensional manifolds with boundaries. After introducing some basic notation
we will explain how one can write the perturbation in superspace in 
a manifestly supersymmetric manner.

\subsection{Superfields and OPEs}

Let us begin by discussing some important aspects of two-dimensional supersymmetric 
field theories. We will work with the standard $\N=(1,1)$ superspace 
(see {\it e.g.}\ \cite{Dorrzapf:1997rx})
\begin{align}
\mathbb{R}^{(2|1,1)}=\{z,\bar{z},\theta,\bar{\theta}\}\ ,\label{StandardSuperspace}
\end{align}
with a boundary along the line $z=\bar{z}$ (for more details see also 
appendix~\ref{App:N1Superspace}). A scalar superfield is of the form 
\begin{align}
\Phi(z,\bar{z},\theta,\bar{\theta})=\phi(z,\bar{z})+\theta\chi(z,\bar{z})
       +\bar{\theta}\bar{\chi}(z,\bar{z})+\theta\bar{\theta}F(z,\bar{z})\ . \label{bulksuperfield}
\end{align}
We denote by $h$ and $\bar{h}$ the (left- and right-moving) 
conformal dimension of the lowest component
of $\Phi$, and by $\Delta=h+\bar{h}$ the total conformal dimension. For the following 
it is important to characterise the behaviour of two such superfields 
as they approach each other. Introducing labels $I,J,\ldots$ to distinguish them,  
the bulk operator product expansion (OPE) takes the form
\begin{align}
\Phi_I&(z_1,\bar{z}_1,\theta_1,\bar{\theta}_1)\, \Phi_J(z_2,\bar{z}_2,\theta_2,\bar{\theta}_2)
\label{BulkBulkOPE} \\
& =\sum_K|z_{12}|^{\Delta_K-\Delta_I-\Delta_J}C_{IJK}^{(1)}
\Bigl[ \Phi_K + \cdots \Bigr] 
+ \sum_L |z_{12}|^{\Delta_{L}-\Delta_I-\Delta_J-1}C_{IJL}^{(2)}
\bigg[ 
\theta_{12}\bar{\theta}_{12} \Phi_L
+\cdots \biggr] \nonumber \\ 
& + \sum_\alpha |z_{12}|^{\Delta_{\alpha}-\Delta_I-\Delta_J-\frac{1}{2}}C_{IJ\alpha}^{(3)} 
\Bigl[ \theta_{12} \Xi_\alpha + \cdots \Bigr]
+ \sum_{\beta} |z_{12}|^{\Delta_{\beta}-\Delta_I-\Delta_J-\frac{1}{2}}C_{IJ\beta}^{(4)} 
\Bigl[\bar\theta_{12} \Xi_\beta + \cdots \Bigr] \ , \nonumber
\end{align}
where $z_{12}$ and $\theta_{12}$ are defined by 
\begin{align}
&z_{12} = \frac{1}{2} (z_1 - z_2) - \theta_1 \theta_2 \ , &&\theta_{12} = \frac{1}{2}(\theta_1-\theta_2) \ ,
\end{align}
and the fields appearing on the right hand side are all evaluated at 
\begin{align}
&\hat{z}_{12}=\frac{1}{2}(z_1+z_2)\  , &&\hat{\theta}_{12}=\frac{1}{2}(\theta_1+\theta_2)\ . 
\end{align}
The ellipses refer to terms that are either of higher order in $|z_{12}|$ or involve 
superdescendants of $\Phi_I$ and $\Xi_\alpha$. Using nomenclature from conformal field 
theory, the first line in (\ref{BulkBulkOPE}) describes the even-even and odd-odd fusion rules, 
respectively, while the second line gives the contribution of the even-odd and odd-even 
fusion rules. The superfields $\Xi_\alpha$ that appear there are fermionic, but cannot 
be written in terms of superdescendants of $\Phi_I$. They will not play a role in the following.

In the presence of a boundary we also need to describe the behaviour as the bulk field 
approaches the boundary. To this end it is convenient to rewrite 
the superfield (\ref{bulksuperfield}) as
\begin{align}
\Phi(x,y,\thp,\thm)=\phi(x,y)+\thp\chi^+(x,y)+\thm\chi^-(x,y)
+\thp\thm G(x,y)\ ,
\label{bulksuperfieldBDY}
\end{align}
where we have introduced real coordinates via $z=x+iy$ and 
$\theta^\pm = (\theta\pm \bar\theta)$, as 
well as 
\begin{align}
&\chi^+=\frac{1}{2}(\chi+\bar{\chi})\ , &&\chi^-=\frac{1}{2}(\chi-\bar{\chi})\ ,
&&G=-\frac{1}{2}F\ .
\end{align}
It is now convenient to expand (\ref{bulksuperfieldBDY}) in powers of the variable 
$\tilde{y}=y-\theta\bar{\theta}$, which becomes small in the vicinity of the boundary.
The most generic expression which can be written down is of the following form
\begin{align}
& \Phi_I(x,y,\thp,\thm) \nonumber \\
& \quad = \sum_iB_{Ii}^{(1)}(2\tilde{y})^{h_i-\Delta_I}
\biggl[\Pi_i(x,\thp)+\cdots \biggr]
+\sum_aB_{Ia}^{(2)}(2\tilde{y})^{h_a-\Delta_I+\frac{1}{2}}
\biggl[\D^+\Psi_a(x,\thp)+\cdots \biggr]\nonumber\\
&\qquad +\thm\left(\sum_j B_{Ij }^{(3)}(2\tilde{y})^{h_j-\Delta_I} 
\biggl[\D^+\Pi_j(x,\thp) + \cdots \biggr]
\right. \nonumber \\
& \qquad \hspace{2cm} \left.
+\sum_b B_{Ib}^{(4)}(2\tilde{y})^{h_b-\Delta_I-\frac{1}{2}}
\biggl[\Psi_b(x,\thp) + \cdots \biggr]\right)
\ ,\label{BdySuperfieldExpansion}
\end{align}
where $\D^+$ is the spinor derivative defined in appendix~A.1 and $B_{Ii}^{(1,3)}$ and $B_{Ia}^{(2,4)}$ 
are some expansion coefficients. $\Pi_i$ and $\Psi_a$ are the most generic boundary superfields which 
can be written using just a single Grassmann variable. They have an expansion as 
\begin{align}
&\Pi_i(x,\thp)=\pi_i(x)+\thp \chi_i(x)\ ,\label{bosonicBDY}\\
&\Psi_a(x,\thp)=\psi_a(x)+\thp \rho_a(x)\ ,\label{fermionicBDY}
\end{align}
where we note that $\Pi_i$ is bosonic, while $\Psi_a$ is fermionic. Finally, $h_i$ and $h_a$ are the 
conformal dimensions of $\pi_i$ and $\psi_a$ respectively. From a superspace point of view, (\ref{BdySuperfieldExpansion}) corresponds to a decomposition of the $\N=(1,1)$ superfield in 
terms of the $\N=1$ superfields (\ref{bosonicBDY}) and (\ref{fermionicBDY}) (and their (super)derivatives). 
In the language of conformal field theory, the terms in the first line correspond to the even fusion rules 
with respect to $\theta^-$ --- the two terms are even and odd with respect to $\theta^+$, 
respectively ---  while the terms in the second and third  line are in the odd fusion channel with respect 
to $\theta^-$. The coefficients can, for example, be determined by analysing the fusion rules, 
using the techniques of appendix~B.2  in \cite{Brunner:2009mn}.

We will also need the OPEs of the boundary superfields with one another, 
in particular
\begin{align}
&\Psi_a(x_1,\thpi{1})\Psi_b(x_2,\thpi{2})=
\sum_iD^{(1)}_{abi}|x_{12}|^{h_i-h_a-h_b}\left[1+(h_i-h_a-h_b)\thpi{12}\hat{\D}^+_{12}\right]\Pi_i(\hat{x}_{12},\hat{\theta}^+_{12})\nonumber\\*
&\hspace{4.2cm}+\sum_cD^{(2)}_{abc}|x_{12}|^{h_c-h_a-h_b-\frac{1}{2}}\left[|x_{12}|\hat{\D}^+_{12}+\thpi{12}\right]\Psi_c(\hat{x}_{12},\hat{\theta}^+_{12})+\cdots\ , \label{PsiPsiOPE}
\end{align}
where we have introduced the variables
\begin{align}
&x_{12}=\frac{1}{2}(x_1-x_2)+\thp_1\thp_2\ , &&\hat{x}_{12}=\frac{1}{2}(x_1+x_2)\ , &&
\theta_{12}^+=\frac{1}{2}(\thp_1-\thp_2)\ , &&\hat{\theta}_{12}^+=\frac{1}{2}(\thp_1+\thp_2)\ ,
\end{align}
along with their corresponding spinor derivative $\hat\D_{12}^+$. From a (boundary) CFT 
point of view the two lines of (\ref{PsiPsiOPE}) again represent the even and odd fusion channel, 
respectively. For later convenience we have also explicitly displayed the first super-descendants, 
which correspond to the terms proportional to $\theta_{12}^+$, along with the appropriate 
numerical factors.

\subsection{Superactions}\label{Sect:SuperspaceCouplings}

With these preparations we can now explain how to formulate manifestly supersymmetric 
perturbations of an $\N=1$ superconformal field theory in the presence of a boundary.

\subsubsection{Bulk deformations}

As is well known, we can write a supersymmetric bulk deformation
as an integral over the standard superspace (\ref{StandardSuperspace})
\begin{align}
S_{\text{bulk}}= \lambda \int d^2z\int d\theta\int d\bar{\theta}\, \,
\Phi(z,\bar{z},\theta,\bar{\theta})\  ,\label{BulkDeformationpre}
\end{align}
where $\lambda$ is the coupling constant corresponding to $\Phi$. In general this deformation 
is however only supersymmetric as long as we consider the theory on manifolds 
without a boundary. Indeed, as explained in \cite{DeWittBook,VoronovBook,RogersBook},
the integral in (\ref{BulkDeformationpre}) is not invariant under generic coordinate 
transformations if the supermanifold over which the integral is taken has a boundary
\cite{Berezin1,BerezinLeites,BernsteinLeites1,BernsteinLeites2,Berezin2}.\footnote{For 
a simple example of this claim see appendix \ref{App:NonInv}.} This is just a reformulation
of the fact that the deformation (\ref{BulkDeformationpre}) breaks supersymmetry in the 
presence of a boundary, as was for example already discovered in \cite{Warner:1995ay}. 

There are several possibilities for how to generalise the Berezin integration to 
supermanifolds with boundaries. For example, one can formulate the integral as a generalised
contour integral \cite{DeWittBook,Rogers1}, or treat the Berezin integral as a differential
operator and introduce a special type of differential form to allow integration over
arbitrary supermanifolds \cite{Rothstein}.
Here we shall follow a different approach \cite{VoronovBook}, and define 
the integral  as the integral over the full superspace, restricted to a certain domain. Let 
$u(x,y,\theta,\bar{\theta})=0$ be the defining equation of the boundary, with 
$u(x,y,\theta,\bar{\theta})<0$ corresponding to the interior. (In particular, the carrier,
{\it i.e.}\ the `bosonic' piece of the superboundary is described by the equation 
$u(x,y,0,0)=0$.) The invariant integral measure of the supermanifold with boundary is then
\begin{align}
\int_{-\infty}^\infty dx \int_{-\infty}^\infty dy\int d^2\theta\, \,
\vartheta\left(-u(x,y,\theta,\bar{\theta})\right)\ .
\end{align}
Here $\vartheta(x)$ is the analytic continuation of the characteristic function, which is defined via its Grassmann expansion. For a boundary function 
\begin{align}
u(x,y,\theta,\bar{\theta})=u_0(x,y)+\theta\bar{\theta}\, u_1(x,y)\ ,\label{GeneraluFunct}
\end{align}
we have the definition
\begin{align}
\vartheta\left(-u(x,y,\theta,\bar{\theta})\right):=\Theta(-u_0(x,y))-\theta\bar{\theta}\, 
u_1(x,y)\, \delta(u_0(x,y))\ , \label{thetaex}
\end{align}
where $\Theta$ and $\delta$ on the right hand side are the usual Heaviside step-function and its first derivative (the Dirac delta-function), respectively.

For the case of the upper half-plane we take the boundary function to be 
\begin{align}
u(x,y,\theta,\bar{\theta})=-\tilde{y}=-y+\theta\bar{\theta}\ .
\end{align}
The choice $u_1(x,y)=1$ is motivated by the requirement that the boundary of the 
supermanifold should have codimension $(1|1)$ as (\ref{FermBdyCond}) demands; 
for example, had we taken $u_1(x,y)=0$, the superspace integral would still be invariant, but 
the boundary would only have co-dimension $(1|0)$. The invariant bulk deformation
(\ref{BulkDeformationpre}) then becomes
\begin{align}
S^{\text{inv}}_{\text{bulk}}&=\lambda
\int  d^2z\int d\theta\int d\bar{\theta}\, \vartheta\!\left(y-\theta\bar{\theta}\right) 
\Phi(z,\bar{z},\theta,\bar{\theta})  \label{BulkDeformation} \\
&=\lambda \left[
\int_{y>0} d^2z\int d\theta\int d\bar{\theta}\, 
\Phi(z,\bar{z},\theta,\bar{\theta})
+ \int_{-\infty}^{\infty} dx \, \int d\theta\int d\bar{\theta}\, \theta\bar\theta\, 
\Phi(x,y=0,\theta,\bar{\theta}) \right]\ , \nonumber
\end{align}
where we have used (\ref{thetaex}).
This result is actually familiar from a field theory point of view, see 
\cite{Warner:1995ay,Hori:2004ja}. Indeed, if we consider the supersymmetry-variation of 
(\ref{BulkDeformationpre}) with respect to (\ref{SuperVariat}), we obtain
\begin{align}
\delta_{\N=1} S_{\text{bulk}}=&
-2  \lambda \int_{y>0} d^2z\int d\theta\int d\bar{\theta}
\left(\epsilon\theta\partial_z-\bar{\epsilon}\bar{\theta}\partial_{\bar{z}}\right) 
\Phi(z,\bar{z},\theta,\bar{\theta})\ .
\end{align}
Using the basis (\ref{boundarybasis1}) and (\ref{boundarybasis2}), as well as the 
boundary conditions (\ref{FermBdyCond}), integration by parts then leads to 
\begin{align}
\delta_{\N=1} S_{\text{bulk}}= \frac{\lambda}{2} 
\int_{-\infty}^\infty dx\int d\thp\epm\Phi(x,y=0,\thp,\thm=0)\, .\label{OffendingBulk}
\end{align}
In general this term is non-vanishing, thus showing that (\ref{BulkDeformationpre}) by itself
is not supersymmetric  \cite{Warner:1995ay}. In order to restore supersymmetry one
therefore has to add to (\ref{BulkDeformationpre}) a pure boundary term whose supervariation
precisely cancels (\ref{OffendingBulk}). This is exactly what the second term in 
(\ref{BulkDeformation}) achieves. 

\subsubsection{Boundary deformations}

For the discussion of the coupled bulk-boundary RG equations we shall also need pure
boundary perturbations. The $(1|1)$ superspace on the boundary does not have
any boundary points itself, and thus the usual superspace measure will be appropriate.
The manifestly supersymmetric boundary deformation is thus of the form
\begin{align}
S_{\text{bdy}}= \mu\, \int dx\int d\thp \, \Psi(x,\thp)\ ,\label{BdyDeformationInvar}
\end{align}
where $\Psi$ is a fermionic superfield, and $\mu$ is the corresponding coupling
constant. Since we have a single fermionic integration in this case, the resulting expression
is bosonic, as must be the case for a term that can be added to the action. Because of 
this reason we cannot write down a similar term involving the bosonic boundary superfield
$\Pi$. The only bosonic term we could write down would be of the form
\begin{align}
S_{\text{bdy}}^{\prime}= \gamma \int dx\int d\thp \thp \, \Pi(x,\thp)\ ,\label{BdyDeformationDef}
\end{align}
but this is  not supersymmetric.

\section{Renormalisation group equations}\label{Sect:RGequations}
\setcounter{equation}{0}

Now we can turn to the analysis of the renormalisation group (RG) equations. We shall
only consider the manifestly supersymmetric deformation terms from above. In particular, we
want to show that the RG equations close among the coupling constants 
$\lambda_I$ and $\mu_a$, corresponding to the manifestly supersymmetric 
deformations (\ref{BulkDeformation}) and (\ref{BdyDeformationInvar}), respectively. 
We will work to leading order in the bulk couplings, but to next-to-leading order in the 
boundary couplings. 

\subsection{The supersymmetric scheme}\label{Sect:N1Regularization}

In the following we shall work in a Wilsonian scheme, which is also sometimes referred to
as the `OPE-scheme' since the coefficients of the leading order terms of the 
RG-equation are proportional to OPE coefficients \cite{Gaberdiel:2008fn}. To derive
the RG equations we consider the expansion of the free energy $e^{\Delta S}$ with
\begin{align}
\Delta S=&\sum_I\lambda_I\int d^2z\int d^2\theta \,
\vartheta\left(y-\theta\bar{\theta}\right)\Phi_I(z,\bar{z},\theta,\bar{\theta})
+\sum_a\mu_a\int dx\int d\thp\, \Psi_a(x,\thp)
\end{align}
in powers of the coupling constants $(\lambda_I,\mu_a)$. Obviously these 
integrals are in general divergent, and we need to regularise them, for example
by introducing a cut-off $\ell$. 
Since the free energy is a physical quantity it should not depend on the value of this cut-off. 
This then requires, as we shall see, that the coupling constants $\lambda_I$ and $\mu_a$ 
are functions of $\ell$. However, in order to be able to re-absorb changes in $\ell$ into a
redefinition of the manifestly supersymmetric terms parametrised by $(\lambda_I,\mu_a)$, 
we need to choose our regulator prescription carefully. In general, the divergencies arise when 
either (i) two bulk fields come close together; (ii) two boundary fields come close together; or
(iii) a bulk field comes close to the boundary. For case (i) and (ii) our regulator will
simply cut off the integrals so that two bulk fields or two boundary fields do not 
come closer than $\ell$. As regards the third divergence, the naive prescription would be
to prevent the bulk field from getting closer than $\tfrac{\ell}{2}$ to the boundary. However,
this prescription would not preserve supersymmetry since we would run into the same
problems as above. We shall therefore use the same idea as there and implement the cut-off
by multiplying the bulk integral by $\vartheta\left(\tilde{y}-\frac{\ell}{2}\right)$. 

As usual we shall make our coupling constants dimensionless by multiplying
them by a suitable power of $\ell$, {\it i.e.}\ by writing the perturbation as 
\begin{align}
\Delta S=&\sum_I\lambda_I \ell^{\Delta_I-1} \int d^2z\int d^2\theta \,
\vartheta\left(y-\theta\bar{\theta}\right)\, 
\vartheta\left(y -\theta\bar{\theta} -\frac{\ell}{2}\right)
\Phi_I(z,\bar{z},\theta,\bar{\theta}) \nonumber \\
& \quad 
+\sum_a\mu_a \ell^{h_a-\frac{1}{2}} \int dx\int d\thp\, \Psi_a(x,\thp) \ .
\end{align}
If we change the cut-off $\ell$ by $\ell \mapsto \ell (1+\delta t)$, then 
from this explicit dependence of the integrals on $\ell$ we get the 
usual leading RG terms
\begin{equation}\label{RG1}
\dot \lambda_I = (1 - \Delta_I) \lambda_I + \cdots \ , \qquad \qquad
\dot \mu_a = (\tfrac{1}{2}- h_a) \mu_a + \cdots \ .
\end{equation}
In addition, we also have a contribution from the {\em implicit} dependence on
$\ell$ via the cut-off prescription. Let us first consider the contribution from the 
first order bulk deformation. Since the dependence on $\ell$ comes from 
the contribution where the bulk field is close to the boundary, we may use
the bulk-boundary OPE (\ref{BdySuperfieldExpansion}) to write 
\begin{align}
A_I &= \int d^2z\int d^2\theta\, \vartheta\left(y-\frac{\ell}{2}-\theta\bar{\theta}\right)
\langle\Phi_I(z,\bar{z},\theta,\bar{\theta})\cdots\rangle\nonumber\\
&=\int d^2z\int d^2\theta\,\vartheta\left(y-\frac{\ell}{2}+\frac{1}{2}\thp\thm\right)
\bigg[\sum_iB_{Ii}^{(1)}(2\tilde{y})^{h_i-\Delta_I}\langle\Pi_i(x,\thp)\cdots\rangle\nonumber\\
&\quad\hspace{1cm}+\sum_{a}B_{Ia}^{(2)}(2\tilde{y})^{h_a-\Delta_I+\frac{1}{2}}
\langle(\D^+\Psi_a)(x,\thp)\cdots\rangle \nonumber\\
&\quad\hspace{1cm}+\thm\sum_aB_{Ia}^{(4)}(2\tilde{y})^{h_a-\Delta_I-\frac{1}{2}}
\langle \Psi_a(x,\thp)\cdots \rangle +\cdots\bigg]\ .\label{A1}
\end{align}
Here we have dropped the ${\cal D}^+\Pi_j$ term since it leads, after 
$\theta^+$  integration, to a total derivative (with respect to $x$), which 
we can ignore. The final ellipses describe terms appearing 
at higher order in $y$. Next we perform the $\thm$ integration to obtain
\begin{align}
A_I &=\int d^2z\int d\thp\thp\sum_iB_{Ii}^{(1)}2^{h_i-\Delta_I}
\left[\Theta\left(y-\frac{\ell}{2}\right)(h_i-\Delta_I)+\delta\left(y-\frac{\ell}{2}\right)y\right] \nonumber\\
&\hspace{7cm}\times\,y^{h_i-\Delta_I-1}\langle\Pi_i(x,\thp)\cdots\rangle\nonumber\\
&\quad+\int d^2z\int d\thp\thp\sum_aB_{Ia}^{(2)}2^{h_a-\Delta_I+\frac{1}{2}}
\left[\Theta\left(y-\frac{\ell}{2}\right)(h_a-\Delta_I+\frac{1}{2})+\delta\left(y-\frac{\ell}{2}\right)y\right] \nonumber\\
&\hspace{7cm}\times\,y^{h_a-\Delta_I-\frac{1}{2}}\langle(\D^+\Psi_a)(x,\thp)\cdots\rangle\nonumber\\
&\quad +\frac{1}{2}\int d^2z\int d\thp\Theta\left(y-\frac{\ell}{2}\right)\sum_aB_{Ia}^{(4)}
(2y)^{h_a-\Delta_I-\frac{1}{2}}\langle \Psi_a(x,\thp)\cdots \rangle+\cdots\ .\label{A11}
\end{align}
In a final step, we can perform the $y$-integration. In the first term, for $h_i\neq \Delta_I$ 
both contributions from the square bracket cancel each other on the boundary 
({\it i.e.}\ for $y\to \ell/2$), leaving just an IR-divergence for 
$y\to \infty$ which we shall, as usual, ignore. 
For $h_i=\Delta_I$ on the other hand, 
the first term of the square bracket is absent, while the second one gives an 
$\ell$-independent contribution which therefore does not contribute to the RG equations. 
A similar argument also applies to the second term. Finally, in the last line
we can simply perform the $y$-integration directly.  The $\ell$-dependence of the 
resulting contribution can then be absorbed into redefining $\mu_a$; more specifically, we 
obtain in this manner the correction to the second equation of (\ref{RG1}) 
\begin{equation}\label{RG2}
\dot \mu_a = (\tfrac{1}{2}- h_a) \mu_a + \frac{1}{2} \sum_I  B_{Ia}^{(4)} \lambda_I 
+ \cdots \ .
\end{equation}
This is the supersymmetric analogue of the bulk-induced source term of 
\cite{Fredenhagen:2006dn}. Note that only a source term for the 
supersymmetric boundary perturbation corresponding to $\Psi_a$ is 
switched on, but not for the supersymmetry breaking perturbation involving
$\Pi_i$. 

\subsection{Higher order contributions}\label{Sect:HigherOrder}

In the following we want to study the quadratic terms in the RG equations
for the boundary coupling constant $\mu_a$. These arise from the implicit
$\ell$-dependence of two types of correlators that we shall discuss in turn.

\subsubsection{The boundary two-point function}

The implicit $\ell$-dependence of the boundary two-point function leads, by the 
usual computation (see for example \cite{Recknagel:2000ri}), to a further correction of 
(\ref{RG2}) 
\begin{equation}\label{RG3}
\dot \mu_a = (\tfrac{1}{2}- h_a) \mu_a + \frac{1}{2} \sum_I  B_{Ia}^{(4)} \lambda_I 
+  \sum_{b,c} D_{abc}^{(2)} \, \mu_b \mu_c  + \cdots \ .
\end{equation}
For the consistency of our manifestly supersymmetric scheme, it is important
that only a correction term corresponding to $\mu_a$ is switched on. This is not obvious
since, on the face of it, we also get a contribution from the first line of (\ref{PsiPsiOPE}),
giving rise to a source term for the non-supersymmetric coupling corresponding
to (\ref{BdyDeformationDef}). The resulting term is of the form
\be\label{quadratic}
\dot \gamma_i \sim  {\cal D}_{iab}\,  \mu_a \mu_b = 
\frac{1}{2} \bigl(  {\cal D}_{iab} +  {\cal D}_{iba} \bigr)\, \mu_a \mu_b \ ,
\ee
where $\gamma_i$ is the coupling constant corresponding to $\Pi_i$ 
in (\ref{BdyDeformationDef}), and 
\begin{align} \label{Diab}
\mathcal{D}_{iab}&=\lim_{\ell\to 0}\ell^{h_a+h_b-h_i+1}\frac{\partial}{\partial\ell}
\int dx\int d\thpi{1}\int d\thpi{2}\int d\thpi{3}\thpi{3}\Theta(|x|-\ell) \nonumber\\
&\hspace{3cm}\times\langle\Psi_a(x,\thpi{1})\Psi_b(0,\thpi{2})\Pi_i ^\ast(\infty,\thpi{3})
\rangle \ ,
\end{align}
with $\Pi_i^\ast$ the conjugate field to $\Pi_i$. 
Using (\ref{PsiPsiOPE}) and performing the $\thpi{1}$- and $\thpi{2}$-integrals,
it is straight-forward to check that this correlator 
is a total derivative in $x$. Another way  to see this is to use methods of 
conformal field theory. After performing the $\theta^+$ integrals
 the integrand of $\mathcal{D}_{iab}$ becomes
\begin{align} 
\mathcal{I}_{iab} & = \langle \pi_i |  
(G_{-1/2} \psi_a) (x) \,  (G_{-1/2} \psi_b) (0) \rangle \nonumber\\ 
&=\langle \pi_i | \Delta_{x,0} (G_{-1/2} )\Bigl[ (G_{-1/2} \psi_a) (x) \,  
\psi_b (0) \Bigr] \rangle- \langle \pi_i | (G_{-1/2} G_{-1/2} \psi_a)(x) \, \psi_b(0) \rangle \nonumber \\
& = - \langle \pi_i | (L_{-1} \psi_a)(x) \, \psi_b(0) \rangle
= - \frac{d}{dx}  \langle \pi_i | \psi_a(x) \, \psi_b(0) \rangle \ , \label{derivative}
\end{align}
where we have used the same notation as in \cite{Brunner:2009mn}. The term
proportional to $\Delta_{x,0}(G_{-1/2})$ vanishes since $\pi_i$ is a highest
weight state (but we do not need to assume that either $\psi_a$ or $\psi_b$
are highest weight). In the final line we have used that $L_{-1}$ is the derivative operator, 
thus implying that the integrand is indeed a total derivative. The integral 
$\mathcal{D}_{iab}$ therefore only gets contributions from $\pm \infty$, 
as well as from $x=\pm \ell$. The former are IR effects which
we can ignore. On the other hand, the contributions from 
$x=\pm \ell$ cancel between $\mathcal{D}_{iab}$ and $\mathcal{D}_{iba}$, and
thus the contribution (\ref{quadratic}) to the RG equation actually vanishes. Thus, at least
to this order, no supersymmetry-breaking term is induced. 

\subsubsection{The bulk boundary correlator}

The other interesting contribution comes from the implicit $\ell$-dependence of the
correlator involving one bulk and one boundary field. In
this case, the  $\ell$-dependence appears only 
in the $\vartheta\left(\tilde{y}-\frac{\ell}{2}\right)$ term of the 
bulk integral. In fact, following the same arguments as in (\ref{A11}), the
only contribution (except for total derivatives, see (\ref{BdySuperfieldExpansion}))
comes again from the final line of (\ref{A11}). Obviously, we have to be 
careful in evaluating the precise coefficient since it now involves the correlation function with the
insertion of an additional boundary field, see \cite{Gaberdiel:2008fn}. In particular, 
we need to worry about the divergence as the boundary field that is switched on by the bulk
field approaches the boundary field in the correlator. As in the discussion in 
section~3.2.1 this will contribute to the RG equation for $\dot\mu_a$. On the face of it, 
it will also give rise to a source term for $\dot\gamma_i$. However, by a 
similar reasoning as in (\ref{derivative}) and  (\ref{BdySuperfieldExpansion}),
it is clear that the corresponding integrand
is a total derivative. Thus the only interesting contribution (apart from IR effects which
we ignore) comes from the term where the bulk induced boundary field is evaluated
on either side of the boundary field. However, these two terms cancel since the
boundary correlator is local, {\it i.e.} independent of the order of the fields. 
(This is a consequence of the fact that one of the two boundary fields comes from
a local bulk field; we also assume that $\psi_a$ does not change the boundary condition, 
as is usually the case for moduli.) Thus again, there is no source term for the 
supersymmetry-breaking coupling (\ref{BdyDeformationDef}), and hence 
the scheme closes (at least to this order) on the supersymmetry-preserving fields. 

The complete RG-equations to this order are then of the form
\begin{align}\label{RG4}
\dot \lambda_I & = (1 - \Delta_I) \lambda_I + \cdots \ , \nonumber \\
\dot \mu_a & = (\tfrac{1}{2}- h_a) \mu_a 
+  \frac{1}{2} \sum_I  B_{Ia}^{(4)} \lambda_I 
+ \sum_{b,c}D_{abc}^{(2)}\, \mu_b \mu_c 
+ \sum_{I,b}\mathcal{E}_{Iab} \, \lambda_I \mu_b + \cdots \ ,
\end{align}
where ${\cal E}_{Iab}$ is given by the integral
(for a similar computation in the purely bosonic case see \cite{Gaberdiel:2008fn})
\begin{align}
\mathcal{E}_{Iab}&=-\lim_{\ell\to 0}\ell^{\Delta_I+h_b-h_a}\int d^2z\int d^2\theta 
\int d\thp_1\int d\thp_2\frac{\partial}{\partial\ell}\vartheta\left(y-\frac{\ell}{2}-\theta\bar{\theta}\right)
\nonumber\\*
&\hspace{2cm}\times\langle\Phi_I(z,\bar{z},\theta,\bar{\theta})\Psi_b(0,\thp_1)
\Psi_a^\ast(\infty,\thp_2)\rangle\nonumber\\*
&\hspace*{-0.1cm} 
-\frac{1}{2}\lim_{\ell\to 0}\sum_c\ell^{h_b+h_c-h_a+\frac{1}{2}}B^{(4)}_{Ic}\int dx
\int d\thp_1\int d\thp_2\int d\thp_3\langle\Psi_c(x,\thpi{1})\Psi_b(0,\thpi{2})
\Psi_a^\ast(\infty,\thp_3)\rangle .\label{Ecoeff}
\end{align}
Here the last line stems from lower order counter-terms and simply subtracts 
the  poles of the first term that would lead to divergencies after integration. 
To evaluate the first line one uses 
\begin{equation}
\frac{\partial}{\partial\ell}\, \vartheta\left(y-\frac{\ell}{2}-\theta\bar{\theta}\right) = 
- \frac{1}{2} \delta\left(y-\frac{l}{2}\right) 
+ \frac{1}{2} \theta\bar\theta\, \delta'\left(y-\frac{l}{2}\right) \ .
\end{equation}


\section{Applications to the $\N=2$ case}
\setcounter{equation}{0}

Next we want to apply these general methods to study the behaviour of D-branes
in string theory. As we have mentioned before, in the context of string theory it is important 
to preserve the $\N=1$ supersymmetry since it is a gauge symmetry. Our manifestly $\N=1$
supersymmetric scheme is therefore the appropriate language for this problem. In particular,
we can use it to re-visit the RG analysis of \cite{Brunner:2009mn} and study how the
results of that paper relate to the matrix factorisation analysis of 
\cite{Baumgartl:2007an}.

\subsection{Obstructions and RG flows from (cc) perturbations}\label{Sec:cc}

In the following we shall study B-type boundary conditions under
(cc) and (ca) bulk perturbations; because of mirror symmetry  this then also covers
the case of A-type branes. As was
explained for example in \cite{Brunner:2009mn}, the perturbation of a B-type brane by
a (ca) deformation is never obstructed, while obstructions can arise in the (cc) case. We shall
therefore concentrate on the (cc) case in the following and come back to the (ca) case
below (see subsection~\ref{ca}). Using our manifestly supersymmetric scheme, a (cc) 
perturbation takes the form
\begin{align}
S^{\text{chiral}}_{\text{bulk}}=\sum_I\lambda_I\int d^2z\int d\thNp
\int d\thNm\vartheta\left(y-\thNp\thNm\right)\Phi^{(cc)}_I(z,\bar{z},\thNp,\thNm)\ .
\label{N2chiralAct}
\end{align}
Here we have used similar conventions as in \cite{Hori:2000ck,Hori} (see also appendix~\ref{App:SuperspaceN2}), and $\Phi^{(cc)}$ is a chiral superfield characterised by the following 
analyticity properties
\begin{align}
\bar{\D}_{(+)}\Phi^{(cc)}_I(z,\bar{z},\thNp,\thNm)=\bar{\D}_{(-)}
\Phi^{(cc)}_I(z,\bar{z},\thNp,\thNm)=0\ ,
\end{align}
with the spinor derivatives $\bar{\D}_{(\pm)}$ given in (\ref{N2SpinorDer}). 

The deformation (\ref{N2chiralAct}) is manifestly $\N=1$ supersymmetric since 
the correction term that was introduced by hand in \cite{Brunner:2009mn} 
(see eq.\ (2.13) of that paper) is now automatically included. We are therefore
in the framework of the previous section, and thus the RG equations (\ref{RG4}) apply. 
To lowest order, the qualitative behaviour of the 
RG flow depends then simply on whether $B_{Ia}^{(4)}$ is non-zero for a marginal field 
$\Psi_a$. 

Actually, it is clear on general grounds that $B_{Ia}^{(4)}\neq 0$ only for irrelevant 
$\Psi_a$. To see this we observe that the coefficient $B_{Ia}^{(4)}$ describes the 
bulk-boundary OPE of the bulk field $\Phi_I$ with a $G$-descendant of the boundary field 
$\psi_a$.  In the (cc) case, the bulk field $\Phi$ has $U(1)$-charges  $q=\bar{q}=1$, 
and thus the $U(1)$-charge of the boundary field in question must at least be $q=1$.
But then its conformal dimension satisfies $h\geq \tfrac{1}{2}$, and the case $h=\tfrac{1}{2}$
is excluded since the $G^+_{-1/2}$ descendant is then a null-vector. Thus no RG flow
is directly switched on, as was already observed in \cite{Brunner:2009mn}. 

On the other hand, the matrix factorisation analysis of  \cite{Baumgartl:2007an} suggests that
the (cc) bulk deformation triggers an RG flow on the boundary. As we have just seen,
to leading order no RG flow is switched on. However, higher order terms may also
lead to an RG flow. In particular, the ${\cal E}  \lambda \mu$ term describes the 
change of conformal dimension of the boundary field corresponding to $\mu$ as 
a consequence of the bulk deformation \cite{Gaberdiel:2008fn}. If this term is positive
for a marginal boundary field $\Psi_a$, the field $\Psi_a$ becomes relevant and thus
triggers an instability of the boundary condition. 
\medskip

In order to see whether this does indeed happen, let us study an explicit example. We
consider the quintic at the Gepner point (for our notation and some useful relations 
see appendix~\ref{App:Gepner}) with the 
tensor product boundary condition corresponding to $L_i=1$
(see \cite{Recknagel:1997sb,Brunner:1999jq}). To be specific, let us
analyse the (cc) perturbation corresponding to the bulk field
\begin{equation}
\phi = (1,-1,0)^{\otimes 5} \otimes \overline{(1,-1,0)^{\otimes 5}} \ ,\label{BulkPrimQuintic}
\end{equation}
where we use the same conventions as in \cite{Brunner:2005fv}. As the 
bulk field is brought to the boundary it can switch on the boundary fields 
given in table~\ref{Tab:BoundaryFieldsCCmoduli} --- in the conventions 
of section~2, these are the lowest components of the superfields appearing in 
(\ref{BdySuperfieldExpansion}). 
\begin{table}[h!]
\begin{center}
\begin{tabular}{|l|c|c|}\hline
\textbf{boundary fields} & $h$ & $q$ \\\hline
\parbox{3.6cm}{\vspace{0.2cm}
$\pi=(2,-2,0)^{\otimes 5}$} & \parbox{0.25cm}{\vspace{0.2cm}$1$} & 
\parbox{0.25cm}{\vspace{0.2cm}$2$} \\
$\psi_{1}=\{(2,-2,0)^{\otimes 4}\otimes (3,3,0)\}$ & $11/10$ & $1$ \\
$\psi_{2}=\{(2,-2,0)^{\otimes 3}\otimes (3,3,2)\otimes (3,3,0)\}$ & $17/10$ & $1$ \\
$\psi_{3}=\{(2,-2,0)^{\otimes 2}\otimes (3,3,2)^{\otimes 2}\otimes (3,3,0)\}$ & $23/10$ & $1$ \\
$\psi_{4}=\{(2,-2,0)\otimes (3,3,2)^{\otimes 3}\otimes (3,3,0)\}$ & $29/10$ & $1$ \\
\parbox{5.4cm}
{$\psi_{5}=\{(3,3,2)^{\otimes 4}\otimes (3,3,0)\}$\vspace{0.2cm}} & 
$7/2$ & $1$ \\\hline
\end{tabular}
\end{center}
\caption{Boundary fields switched on by the $(cc)$ bulk modulus $\phi$ 
of (\ref{BulkPrimQuintic}).}
\label{Tab:BoundaryFieldsCCmoduli}
\end{table}
Here curly brackets denote all possible permutations of the five factors, and
we have used that 
\begin{equation}
(1,-1,0) \otimes (1,-1,0) = (3,3,2) \oplus (2,-2,0) \ .
\end{equation}
Note that $(2,-2,0)$ is a primary field, while $(3,3,2)$ is a $G$-descendant of 
$(3,3,0)$. 
The field $\pi$ is bosonic, and is in fact precisely the boundary field that is present 
in the manifestly $\N=1$ supersymmetric formulation, 
see (\ref{BulkDeformation}). The other fields
$\psi_{a}$, $a=1,\ldots,5$, are the lowest components of fermionic superfields 
$\Psi_a$. As explained above on general grounds, they have indeed
$h>1/2$ and thus lead to irrelevant perturbations.

The matrix factorisation analysis of \cite{Baumgartl:2007an} suggests that the 
boundary modulus corresponding to $\pi$ should be switched on by the bulk perturbation.
The corresponding modulus field is obtained by applying a full unit  
of spectral flow $\mathcal{S}$ to the field $\pi$
\begin{align}
\hat{\psi}=\mathcal{S}\pi = (1,1,0)^{\otimes 5} \ .
\end{align}
The field $\hat\psi$ has indeed $h=\tfrac{1}{2}$ and $q=-1$, and it is the lowest component of 
a fermionic boundary superfield $\widehat\Psi$. Based on the matrix factorisation analysis we
would therefore expect that this field becomes tachyonic as a consequence of the bulk
perturbation. To see whether this is the case we need to study the correlator
\begin{equation}
{\mathcal E} \sim 
\langle 
(G^-_{-1/2} \bar{G}^-_{-1/2} \phi) \, (G^+_{-1/2} \hat\psi)\,  (G_{-1/2} \psi_b)^\ast\rangle \ ,
\end{equation}
where $\psi_b$ is a marginal boundary field. Actually, as explained just before 
(\ref{A11}), only the channel  where the bulk field switches on a $G$-descendant 
of one of the $\psi_a$ boundary fields contributes to ${\cal E}$. 
Thus the relevant correlator is
\begin{equation}
{\mathcal E}\sim 
\langle (G^-_{-1/2} \psi_a) \, (G^+_{-1/2} \hat\psi)\,  (G_{-1/2} \psi_b)^\ast \rangle \ .
\end{equation}
Using (\ref{FusionN2minimal}) one can indeed show that the correlator is only non-zero if  
$\psi_b = \hat\psi^\ast$, leading to the RG equation
\be
\dot\mu^\ast = {\cal E} \, \lambda \, \mu \ ,
\ee
where $\mu^\ast$ is the coupling constant for $\hat\Psi^\ast$, while $\lambda$ corresponds to 
the bulk deformation (\ref{BulkPrimQuintic}). Since the bulk perturbation must be real, 
it must also involve $\Phi^\ast$, and this leads to the RG term
\be
\dot\mu = {\cal E} \, \lambda \, \mu^\ast \ ,
\ee
where $\mu$ is the coupling constant for $\hat\Psi$. Taking these two equations together 
it is then clear that the conformal dimension of the fields
$\hat{\psi}_\pm=\hat{\psi}\pm\hat{\psi}^\ast$ is shifted 
by $\pm {\cal E}\lambda$. Irrespective of the sign of $\mathcal{E}$, one of
the two fields  therefore becomes relevant and thus triggers an instability. At 
least qualitatively, the corresponding flow should be the flow 
predicted in \cite{Baumgartl:2007an} from the matrix factorisation point of view. 
A detailed comparison is, however, difficult because it is not clear how to 
identify the RG scheme from the matrix factorisation analysis.

We have also checked this conclusion for other perturbations of other 
tensor product branes, and the situation is always exactly as above. On the other
hand, the analysis is different for the permutation brane case of \cite{Baumgartl:2007an}  
since at the permutation point the effective superpotential has a zero of higher order. 
In terms of the above RG analysis this translates to the statement that the
${\cal E}$ coefficient is zero, and that only a higher order correlator, involving
a larger number of boundary moduli, is non-zero. This therefore agrees again 
nicely with the expectations from \cite{Baumgartl:2007an}.

\subsection{Manifestly $\N=2$ description of (ca) perturbations}\label{ca}

For the case of a (ca) deformation, not only the $\N=1$ supersymmetry can always be
preserved (by adding suitable boundary terms to the action), but also 
the $\N=2$ symmetry \cite{Brunner:2009mn}.  This suggests that the (ca) case should 
allow for a manifestly $\N=2$ formulation. Using similar conventions as in \cite{Hori:2000ck,Hori} 
(see also appendix~\ref{App:SuperspaceN2}) the (ca) deformation can indeed be written as 
\begin{align}
&S^{\text{twist}}_{\text{inv}}=\sum_I\lambda_I
\int d^2z\int d\thNp\int d\bthNm\vartheta\left(y-\thNp\bthNm\right)
\Phi^{(ca)}_I(z,\bar{z},\thNp,\bthNm)\ .\label{N2twistActInv}
\end{align}
Here $\Phi^{(ca)}_I(z,\bar{z},\thNp,\bthNm)$ is a twisted chiral superfield which 
satisfies the following analyticity properties
\begin{align}
\bar{\D}_{(+)}\Phi_I^{(ca)}(z,\bar{z},\thNp,\bthNm)=\D_{(-)}\Phi_I^{(ca)}(z,\bar{z},\thNp,\bthNm)=0\ .
\end{align}
Here $\bar{\D}_{(+)}$ and $\D_{(-)}$ are two of the $\N=2$ spinor derivatives which are defined in (\ref{N2SpinorDer}).

The deformation (\ref{N2twistActInv}) preserves the full $\N=2$ supersymmetry since the 
supervariation (\ref{N2variation}) leads to 
\begin{align}
&\delta_{\N=2} S^{\text{twist}}_{\text{inv}}\nonumber\\
&=2i\sum_I \lambda_I \int d^2z\int d\thNp\int d\bthNm\left[
\epsilon_{(+)}\bthNm\partial_--\bar{\epsilon}_{(-)}\thNp\partial_+\right]
\Phi^{(ca)}_I(z,\bar{z},\thNp,\bthNm)\nonumber\\
&+2i\sum_I \lambda_I \int d^2z\int d\thNp\int d\bthNm\delta\left(y\right)
\left[\epsilon_{(+)}\bthNm+\thNp\bar{\epsilon}_{(-)}\right]
\Phi^{(ca)}_I(z,\bar{z},\thNp,\bthNm)\ ,\label{TwistSUSYvar} 
\end{align}
which vanishes upon partial integration. 

One may wonder how lines of marginal stability appear in this formulation. Expanding
out the $\vartheta$ function in (\ref{N2twistActInv}), we can write $S^{\text{twist}}_{\text{inv}}$ as 
\begin{equation}
S^{\text{twist}}_{\text{inv}}=\sum_I\lambda_I \int d^2z\int 
d\thNp\int d\bthNm\left[\Theta(y)-\thNp\bthNm\delta(y)\right]\Phi^{(ca)}_I(z,\bar{z},\thNp,\bthNm)\ .
\nonumber
\end{equation}
The second term in the bracket describes a boundary term that needs to be switched on in
order to preserve the ${\cal N}=2$ supersymmetry. This boundary term can be written as 
\begin{equation}\label{N2bound}
-\sum_{i,I}B_{Ii}\lambda_I \int d^2z\int d\thNp\int d\bthNm\thNp\bthNm\delta(y)
(2y)^{h_i-\Delta_I}\Pi_i^{(ca)}(x,\thNp,\bthNm)+\cdots\ ,
\end{equation}
where we have used the bulk boundary OPE which in the $\N=2$ context takes the form 
\begin{equation}
\Phi^{(ca)}_I(z,\bar{z},\thNp,\bthNm)=\sum_iB_{Ii}(2\tilde{y})^{h_i-\Delta_I}
\Pi_i^{(ca)}(x,\thNp,\bthNm)\ .\label{CAbulkBoundary}
\end{equation}
Here $\Pi^{(ca)}$ is a bosonic boundary multiplet, which depends on both $\thNp$ and 
$\bthNm$. It is clear from (\ref{N2bound}) that the additional boundary term is only 
well-defined for $h_i\geq \Delta_I$ but is divergent otherwise. In particular, if we consider a 
deformation by a bulk modulus ($\Delta_I=1$) the boundary correction is only well-defined if 
$h_i\geq 1$, {\it i.e.}\ if the boundary fields switched on by $\Phi$ are marginal or irrelevant. 

Starting from a point in moduli space where all $h_i\geq 1$, we reach a line of marginal 
stability as one of them becomes relevant \cite{Brunner:2009mn}. At this point the manifestly 
supersymmetric scheme ceases to be well-defined and becomes rather formal. Thus the above 
manifestly ${\cal N}=2$ supersymmetric description is not in conflict with the existence of lines of
marginal stability.


\section{Conclusions}\label{Sect:Conclusions}

In this paper we have shown that there exists a manifestly $\N=1$ supersymmetric
RG scheme (at least up to next-to-leading order) for the coupled problem of $\N=1$ preserving 
bulk and boundary perturbations. Since the $\N=1$ superconformal symmetry is a gauge 
symmetry of superstring theory, this is the appropriate scheme in this context.
We have applied our results to the study of B-type branes under (cc) deformations. 
In particular, we have shown that the bulk-induced source for relevant and marginal boundary
perturbations always vanishes in this case, even if the brane is obstructed. The obstruction
manifests itself rather in that a boundary modulus becomes tachyonic, thus triggering
an RG-flow in the corresponding direction in moduli space. We have also seen 
that our results agree, at least qualitatively, with the predictions of \cite{Baumgartl:2007an}. 
A quantitative comparison is problematic since it is not clear how to identify the 
RG scheme from the matrix factorisation analysis.

For (ca) perturbations of B-type branes, on the other hand, no obstructions are believed to 
appear (see for example \cite{Brunner:2009mn}). This is reflected in the fact that a 
manifestly $\N=2$ supersymmetric RG scheme exists in this case (see section~\ref{ca}).
Lines of marginal stability manifest themselves from this point of view as a
breakdown of this scheme. It would be interesting to study this more explicitly in
examples, and see whether this perspective can shed any light
on the wall-crossing formulae of $\N=2$ theories, see for example 
\cite{Gaiotto:2008cd,Gaiotto:2009hg}.


\section*{Acknowledgements}
We thank in particular Cornelius Schmidt-Colinet for many useful
discussions and collaboration at an early stage. We also acknowledge
useful conversations with Ilka Brunner and Christoph Keller.
This work is supported by the Swiss National Science Foundation.


\appendix
\renewcommand{\theequation}{\Alph{section}.\arabic{equation}}
\setcounter{equation}{0}
\section{Superspace conventions}
In this appendix we outline our conventions for the $\N=(1,1)$ and $\N=(2,2)$ 
superspace which we will use throughout this work.


\subsection{$\N=(1,1)$ conventions}\label{App:N1Superspace}
Let us begin by describing the standard $\N=(1,1)$ superspace, which is spanned by the 
coordinates of (\ref{StandardSuperspace}). We will first discuss our notation for the 
bulk, and then introduce a boundary along the line $z=\bar{z}$.

\subsubsection{Bulk superspace}
The supercharges corresponding to the coordinates given in (\ref{StandardSuperspace}) read
\begin{align}
&\Q=\frac{\partial}{\partial\theta}-\theta\partial_z\ , &&\text{and} &&
\bar{\Q}=\frac{\partial}{\partial\bar{\theta}}-\bar{\theta}\partial_{\bar{z}}\ ,
\label{N1BulkSupercharges}
\end{align}
which are combined with constant spinors $\epsilon$ and $\bar{\epsilon}$ to give the following supervariation
\begin{align}
\delta_{\N=1}=\epsilon\Q-\bar{\epsilon}\bar{\Q}\  .\label{SuperVariat}
\end{align}
For later convenience we also introduce the covariant derivatives
\begin{align}
&\D=\frac{\partial}{\partial\theta}+\theta\partial_z\ , &&\text{and} &&
\bar{\D}=\frac{\partial}{\partial\bar{\theta}}+\bar{\theta}\partial_{\bar{z}}\ ,
\end{align}
which satisfy the important relation
\begin{align}
&\D^2=\partial_z\,, &&\text{and} &&\bar{\D}^2=\partial_{\bar{z}}\ .
\end{align}
Moreover, in view of dealing with superconformal field theories, we 
mention that the conformal dimensions of the bosonic and fermionic variables are 
\begin{align}
&h_z=h_{\bar{z}}=-1\ ,&&\text{and} &&h_{\theta}=h_{\bar{\theta}}=-\frac{1}{2}\  ,
\end{align}
which, in particular, implies that the conformal dimension of the integral measure 
of the superspace (\ref{StandardSuperspace}) is 
\begin{align}
h_{\int dz\int d\bar{z}\int d\theta \int d\bar{\theta}}=-1\  .
\end{align}


\subsubsection{Boundary superspace}

Next we introduce a boundary along the line $z=\bar{z}$. For most of the computations 
in the main body of this work it is much more convenient to switch to a 
real basis for the bosonic coordinates. More precisely we introduce
\begin{align}
&z=x+iy\ , &&\text{and} &&\theta=\frac{1}{2}(\thp+\thm)\ , 
&&\text{and} &&\epsilon=\frac{1}{2}(\epp+\epm)\ ,\label{boundarybasis1}\\
&\bar{z}=x-iy\ , &&\text{and} &&\bar{\theta}=\frac{1}{2}(\thp-\thm)\ , 
&&\text{and} &&\bar{\epsilon}=\frac{1}{2}(\epp-\epm)\ .\label{boundarybasis2}
\end{align}
In this basis the boundary is given by the line $y=0$. At this locus only the 
sum of the bulk supercharges (\ref{N1BulkSupercharges}) will remain 
unbroken\footnote{Strictly speaking the most generic boundary condition would be 
$\Q-e^{2\pi i\eta} \left. \bar{\Q} \right|_{y=0}=0$ for an arbitrary phase $\eta$. Since this phase, 
however, will not play any role in our computations we simply choose $\eta=0$.}, 
which entails for the Grassmann variables the following trivial boundary condition
\begin{equation}
\left. \theta=\bar{\theta} \right|_{y=0}\ , \quad \hbox{{\it i.e.}}\quad
\left. \thm=0 \right|_{y=0}\ , \qquad \text{and} \qquad 
\left. \epsilon=-\bar{\epsilon} \right|_{y=0}\ , \quad \hbox{{\it i.e.}}\quad
\left. \epp=0\right|_{y=0}\ .\label{FermBdyCond}
\end{equation}
Thus we can view the boundary as a one-dimensional superspace spanned by the variables
\begin{align}
\mathbb{R}^{(1|1)}=\{x,\thp\}\ .\label{boundarySuperSpace}
\end{align}
For completeness, we also introduce the corresponding supercharge and spinor derivative
\begin{align}
&\Q^+=\frac{\partial}{\partial\thp}-\thp\frac{\partial}{\partial x}\ ,
&&\text{and} &&
\D^+=\frac{\partial}{\partial\thp}+\thp\frac{\partial}{\partial x}\ .\label{SpinDerBound}
\end{align}


\subsection{$\N=(2,2)$ conventions}\label{App:SuperspaceN2}

Similar to the $\N=(1,1)$ superspace of the previous section we will now also 
outline our notations for the $\N=(2,2)$ standard superspace, which will be 
relevant for section \ref{ca}. We  begin again with the bulk 
superspace, and then introduce a boundary at $z=\bar{z}$.

\subsubsection{Bulk superspace}

We shall use standard $\N=(2,2)$ superspace, which we will split in 
two light-cone sectors as 
\begin{align}
\mathbb{R}^{(2|2,2)}=\mathbb{R}_L^{(1|2)}\times \mathbb{R}_R^{(1|2)}=
\{x_+,\thNp,\bthNp\}\times \{x_-,\thNm,\bthNm\}\ .\label{N2standardSuper}
\end{align}
Translations in the Grassmann directions ({\it i.e.}\ $\N=2$ supersymmetry 
transformations) are generated by the supercharges 
\begin{align}
\Q_{(\pm)} =\frac{\partial}{ \partial \thNpm} + i \bthNpm
\partial_\pm\ , 
&& &\bar\Q_{(\pm)}  = - \frac{\partial}{ \partial \bthNpm} 
- i \thNpm  \partial_\pm \ , \label{N2Supercharges}
\end{align}
where $\partial_\pm = \frac{1}{2} (\partial_0 \pm \partial_1)$. The only 
non-vanishing anti-commutators of these generators are 
\begin{align}
\{ \Q_{(\pm)}  , \bar\Q_{(\pm)}  \} = - 2 i \partial_\pm \ .
\end{align}
For completeness, we also introduce the corresponding spinor derivatives, 
which take the form
\begin{align}
&\D_{(\pm)}=\frac{\partial}{ \partial \thNpm} - i \bthNpm \partial_\pm\ , 
&& \bar\D_{(\pm)} = - \frac{\partial}{ \partial \bthNpm} 
+ i \thNpm  \partial_\pm \ , \label{N2SpinorDer}
\end{align}
and which anti-commute with all the $\Q_{(\pm)} $ and $\bar\Q_{(\pm)}$, 
and have only the following non-trivial  anti-commutator relations
\begin{align}
\{ \D_{(\pm)} , \bar\D_{(\pm)} \} =  2 i \partial_\pm \ . 
\end{align}
Introducing the constant spinors $\epsilon_{(\pm)}$ and $\bar{\epsilon}_{(\pm)}$ we can 
parametrise the supervariation as
\begin{align}
\delta_{\N=2}=\epsilon_{(+)}\Q_{(-)}-\epsilon_{(-)}\Q_{(+)}-\bar{\epsilon}_{(+)}\bar{Q}_{(-)}+\bar{\epsilon}_{(-)}\bar{Q}_{(+)}\,.\label{N2variation}
\end{align}

\subsubsection{Boundary superspace}
We will now introduce a boundary in the bosonic coordinates. In the 
basis of (\ref{N2standardSuper}) we choose the line $x_+=x_-$. Just as in 
the $\N=(1,1)$ case, only half of the four supercharges (\ref{N2Supercharges}) are 
preserved along this line. In fact, ignoring irrelevant phase factors, there 
are two distinct boundary conditions
\begin{align}
&\text{A-type:}&&\epsilon\equiv \epsilon_{(+)}=\bar{\epsilon}_{(-)}\ ,&&\text{and} 
&&\bar{\epsilon}=\bar{\epsilon}_{(+)}=\epsilon_{(-)}\ ,\label{N2Atype}\\
&\text{B-type:}&&\epsilon\equiv \epsilon_{(+)}=-\epsilon_{(-)}\ ,&&\text{and} &&\bar{\epsilon}=\bar{\epsilon}_{(+)}=-\bar{\epsilon}_{(-)}\ .\label{N2Btype}
\end{align}
Throughout this work we will just consider B-type boundary conditions; this is not a restriction 
since we may use mirror symmetry to obtain the corresponding statements for 
A-type boundary conditions.  

\section{Generic superspace coordinate transformations}\label{App:NonInv}
\setcounter{equation}{0}

In this appendix we illustrate, with a simple example,  the fact that the usual 
Berezin integration fails to be invariant in the presence of a boundary of the carrier 
manifold. Let us consider an arbitrary function $F(x,y,\theta,\bar{\theta})$ which lives 
on the superspace (\ref{StandardSuperspace}) and which has the following Grassmann 
expansion
\begin{align}
F(x,y,\theta,\bar{\theta})=F_{(0,0)}(x,y)+\theta F_{(1,0)}(x,y)+\bar{\theta}F_{(0,1)}(x,y)+\theta\bar{\theta}F_{(1,1)}(x,y)\  .
\end{align}
Let us consider an integral of this function over the supermanifold
(\ref{StandardSuperspace}), where we have a boundary at the line $y=0$. 
A typical integral using the naive integral prescription for compact supermanifolds is for 
example given by
\begin{align}
\mathcal{I}^{\text{Ber}}&=\int_{-\infty}^\infty dx\int_0^\infty dy \int d\bar{\theta}\int d\theta \,
F(x,y,\theta,\bar{\theta})
=\int_{-\infty}^\infty dx\int_0^\infty dy \, F_{(1,1)}(x,y) \ .\label{UnTransformed}
\end{align}
Now suppose we make the coordinate transformation
\begin{align}
y\mapsto\tilde{y}=y-\theta\bar{\theta} \label{ExampCoordTrans}
\end{align}
before the Grassmann integration. By the naive Berezin rules we would find
\begin{align}
\mathcal{I}^{\text{Ber}}&=\int_{-\infty}^\infty dx\int_0^\infty d\tilde{y} 
\int d\bar{\theta}\int d\theta \, F(x,\tilde{y}+\theta\bar{\theta},\theta,\bar{\theta})\nonumber\\
&=\int_{-\infty}^\infty dx\int_0^\infty d\tilde{y} 
\left(F_{(1,1)}(x,\tilde{y})+\partial_{\tilde{y}}F_{(0,0)}(x,\tilde{y})\right)\nonumber\\
&=\int_{-\infty}^\infty dx\int_0^\infty d\tilde{y}\, 
F_{(1,1)}(x,\tilde{y})-\int_{-\infty}^\infty dxF_{(0,0)}(x,0)\ ,
\end{align}
which differs from (\ref{UnTransformed}) by an additional integral over the boundary 
of the carrier ({\it i.e.}\ an integral along the line $y=0$). The reason for this discrepancy 
is that the Berezin transformation rules have only instructed us to transform the integrand, 
but they fail to also adapt the boundaries of the bosonic $y$ integration. 

Let us therefore consider a manifestly invariant integral. In order to make contact with 
section~\ref{Sect:SuperspaceCouplings} we choose an integral with a boundary function 
$u(x,y,\theta,\bar{\theta})=-y+\theta\bar{\theta}$
\begin{align}
\mathcal{I}^{\text{inv}}&=\int_{-\infty}^\infty dx\int_{-\infty}^\infty dy 
\int d\bar{\theta}\int d\theta\, \vartheta(y-\theta\bar{\theta}) F(x,y,\theta,\bar{\theta})\nonumber\\
&=\int_{-\infty}^\infty dx\int_0^\infty dy \, F_{(1,1)}(x,y)-\int_{-\infty}^\infty dx\, F_{(0,0)}(x,0)\ ,\label{ExInvPure}
\end{align}
where we have used the expansion (\ref{thetaex}). Now let us again study this integral after the 
coordinate transformation (\ref{ExampCoordTrans}) 
\begin{align}
\mathcal{I}^{\text{inv}}&=
\int_{-\infty}^\infty dx\int_{-\infty}^\infty d\tilde{y} \int d\bar{\theta}
\int d\theta\, \Theta(\tilde{y}) \, 
F(x,\tilde{y}+\theta\bar{\theta},\theta,\bar{\theta})\ .\label{ExInvTrans}
\end{align}
As we can see, there 
is no need to change the integration range of the $\tilde{y}$ variable, since
it is anyway unbounded. Calculating the Grassmann integrals in (\ref{ExInvTrans}) we 
then obtain
\begin{align}
\mathcal{I}^{\text{inv}}&=
\int_{-\infty}^\infty dx\int_{-\infty}^\infty d\tilde{y}\int d\bar{\theta}\int d\theta\,\Theta(\tilde{y})
\left( F(x,\tilde{y},\theta,\bar{\theta})+\theta\bar{\theta}\,\partial_{\tilde{y}}\,
F(x,\tilde{y},\theta,\bar{\theta})\right)\nonumber\\
&=\int_{-\infty}^\infty dx\int_{0}^\infty d\tilde{y} \, F_{(1,1)}(x,\tilde{y})-\int_{-\infty}^\infty dx\, 
F_{(0,0)}(x,0) \ ,
\end{align}
which indeed agrees with (\ref{ExInvPure}). With this modified prescription
the superspace integral is thus invariant under generic coordinate transformations.

\setcounter{equation}{0}
\section{Gepner Point Description}\label{App:Gepner}
In view of the particular example studied in section~\ref{Sec:cc} we compile
here some basic notations and useful relations. At the Gepner point, the quintic
Calabi-Yau can be described as a $\mathbb{Z}_5$ orbifold of a five-fold product of 
$\mathcal{N}=2$ minimal models, each with $k=3$. The central charge of a single minimal 
model at level $k$ is given by
\begin{align}
c=\frac{3k}{k+2}\ ,\label{MinModCentralCharge}
\end{align}
and thus the total charge of five copies with $k=3$ gives
$c_{\text{tot}}=9$, as is appropriate for a Calabi-Yau manifold. 

The representations  $\mathcal{H}_{(l,m,s)}$ of a single minimal model
are labelled by triples of integers $(l,m,s)$, where $l=0,1,\ldots,k$, while $m$ and $s$ are 
defined modulo $2k+4$ and $4$, respectively. All three labels have to sum up to an even 
integer
\begin{align}
l+m+s=0\,\,\text{mod}\,2\ ,
\end{align}
and we have the {\em field identification}
\begin{align}
(l,m,s)\sim(k-l,m+k+2,s+2)\ .\label{FieldIdentification}
\end{align}
States with $s$ even belong to the Neveu-Schwarz (NS)-sector, while states with 
$s$ odd live in the Ramond (R)-sector. If $|m-s|\leq l$ and $s\in\{-1,0,1,2\}$ the conformal 
weight and $U(1)$ charge of the ground state is given by
\begin{align}\label{eigenv}
h(l,m,s)=\frac{l(l+2)-m^2}{4(k+2)}+\frac{s^2}{8}\ ,&&\text{and} &&
q(l,m,s)=\frac{s}{2}-\frac{m}{k+2}\ .
\end{align}
Finally, the fusion rules are simply described by 
\begin{equation}\label{FusionN2minimal}
(l_1,m_1,s_1) \otimes (l_2,m_2,s_2) = 
\sum_{l=|l_1-l_2|}^{\min(l_1+l_2,2k-l_1-l_2)} \, (l,m_1+m_2,s_1+s_2) \ ,
\end{equation}
where the sum over $l$ is over every second $l$. 

The full state-space at the Gepner point is then spanned by
\begin{align}
\bigotimes_{i=1}^5\mathcal{H}_{(l_i,m_i+n,s_i)}\otimes 
\bar{\mathcal{H}}_{(l_i,m_i-n,\bar{s}_i)}\ ,\label{QuniticStateSpace}
\end{align}
where $\mathcal{H}_{(l_i,m_i,s_i)}$ denotes the $(l_i,m_i,s_i)$ representation in the 
$i$-th minimal model, and $n=0,1,\ldots,4$ describes the twist sectors of the 
$\mathbb{Z}_5$ orbifold. $s_i$ and $\bar{s}_i$ are either all odd (R-sector) or all 
even (NS-sector).  Finally, the (cc) fields take the general form
\begin{align}
&\Phi^{(cc)}_{l_1,l_2,l_3,l_4,l_5}=\prod_{i=1}^5(l_i,-l_i,0)\otimes \overline{(l_i,-l_i,0)}
\ ,
\end{align}
where $l_i\leq k_i=3$. It follows from (\ref{eigenv}) that these states indeed have
$h=\tfrac{q}{2}$ and $\bar{h}=\tfrac{\bar{q}}{2}$. 



\begin{thebibliography}{99}

\bibitem{Denef:2007vg}
F.~Denef and G.W.~Moore,
{\it Split states, entropy enigmas, holes and halos},
{\sf arXiv:hep-th/0702146}.

\bibitem{Gaiotto:2008cd}
D.~Gaiotto, G.W.~Moore and A.~Neitzke,
{\it Four-dimensional wall-crossing via three-dimensional field theory},
{\sf arXiv:0807.4723 [hep-th]}.

\bibitem{Gaiotto:2009hg}
D.~Gaiotto, G.W.~Moore and A.~Neitzke,
{\it Wall-crossing, Hitchin systems, and the WKB approximation},
{\sf arXiv:0907.3987 [hep-th]}.

\bibitem{Brunner:2009mn} 
I.~Brunner, M.R.~Gaberdiel, S.~Hohenegger and C.A.~Keller, 
{\it Obstructions and lines of marginal stability from the world-sheet}, 
JHEP {\bf 0509} (2009) 007; 
{\sf arXiv:0902.3177 [hep-th]}. 

\bibitem{Fredenhagen:2006dn} 
S.~Fredenhagen, M.R.~Gaberdiel and C.A.~Keller, 
{\it Bulk induced boundary perturbations}, 
J.\ Phys.\ A  {\bf 40} (2007) F17;
{\sf arXiv:hep-th/0609034}.

\bibitem{Fredenhagen:2007rx}
S.~Fredenhagen, M.R.~Gaberdiel and C.~A.~Keller,
{\it Symmetries of perturbed conformal field theories},
J.\ Phys.\ A  {\bf 40} (2007) 13685; 
{\sf arXiv:0707.2511 [hep-th]}.

\bibitem{Warner:1995ay} N.P.~Warner,
{\it Supersymmetry in boundary integrable models},
Nucl.\ Phys.\  B {\bf 450} (1995) 663;
{\sf arXiv:hep-th/9506064}.

\bibitem{Hori:2004ja} 
K.~Hori and J.~Walcher, 
{\it F-term equations near Gepner points}, 
JHEP {\bf 0501} (2005) 008;
{\sf arXiv:hep-th/0404196}.

\bibitem{Gaberdiel:2008fn} 
M.R.~Gaberdiel, A.~Konechny and C.~Schmidt-Colinet, 
{\it Conformal perturbation theory beyond the leading order}, 
J.\ Phys.\ A  {\bf 42} (2009) 105402; 
{\sf arXiv:0811.3149 [hep-th]}.

\bibitem{Baumgartl:2007an} 
M.~Baumgartl, I.~Brunner and M.R.~Gaberdiel, 
{\it D-brane superpotentials and RG flows on the quintic}, 
JHEP {\bf 0707} (2007) 061; {\sf arXiv:0704.2666 [hep-th]}.

\bibitem{Baumgartl:2008qp}
M.~Baumgartl and S.~Wood,
{\it Moduli webs and superpotentials for five-branes},
JHEP {\bf 0906} (2009) 052; 
{\sf arXiv:0812.3397 [hep-th]}.

\bibitem{Jockers:2008pe}
H.~Jockers and M.~Soroush, 
{\it Effective superpotentials for compact D5-brane 
Calabi-Yau geometries}, 
Commun.\ Math.\ Phys.\  {\bf 290} (2009) 249; {\sf arXiv:0808.0761 [hep-th]}.

\bibitem{Alim:2009rf}
M.~Alim, M.~Hecht, P.~Mayr and A.~Mertens, 
{\it Mirror symmetry for toric branes on compact hypersurfaces,} 
JHEP {\bf 0909} (2009) 126; 
{\sf arXiv:0901.2937 [hep-th]}.

\bibitem{Alim:2009bx}
M.~Alim, M.~Hecht, H.~Jockers, P.~Mayr, A.~Mertens and M.~Soroush, 
{\it Hints for off-shell mirror symmetry in type II/F-theory compactifications}, 
{\sf arXiv:0909.1842 [hep-th]}.

\bibitem{Grimm:2009ef}
T.W.~Grimm, T.W.~Ha, A.~Klemm and D.~Klevers,
{\it Computing brane and flux superpotentials in F-theory compactifications},
{\sf arXiv:0909.2025 [hep-th]}.

\bibitem{Aganagic:2009jq}
M.~Aganagic and C.~Beem,
{\it The geometry of D-brane superpotentials},
{\sf arXiv:0909.2245 [hep-th]}.

\bibitem{Ooguri:1996ck}
H.~Ooguri, Y.~Oz and Z.~Yin,
{\it D-branes on Calabi-Yau spaces and their mirrors},
Nucl.\ Phys.\  B {\bf 477} (1996) 407;
{\sf arXiv:hep-th/9606112}.

\bibitem{Hanany:1997vm}
A.~Hanany and K.~Hori,
{\it Branes and N = 2 theories in two dimensions},
Nucl.\ Phys.\  B {\bf 513} (1998) 119;
{\sf arXiv:hep-th/9707192}.

\bibitem{Hori:2000ck} K.~Hori, A.~Iqbal and C.~Vafa, 
{\it D-branes and mirror symmetry}, 
{\sf  arXiv:hep-th/0005247}.

\bibitem{Hori} K.~Hori,
{\it Linear models of supersymmetric D-branes}, 
{\sf arXiv:hep-th/0012179}.

\bibitem{Albertsson:2001dv}
C.~Albertsson, U.~Lindstr\"om and M.~Zabzine,
{\it N = 1 supersymmetric sigma model with boundaries. I},
Commun.\ Math.\ Phys.\  {\bf 233} (2003) 403;
{\sf arXiv:hep-th/0111161}.
  
\bibitem{Albertsson:2002qc}
C.~Albertsson, U.~Lindstr\"om and M.~Zabzine,
{\it N = 1 supersymmetric sigma model with boundaries. II},
Nucl.\ Phys.\  B {\bf 678} (2004) 295;
{\sf arXiv:hep-th/0202069}.

\bibitem{Lindstrom:2002jb}
U.~Lindstr\"om and M.~Zabzine,
{\it N = 2 boundary conditions for non-linear sigma models and Landau-Ginzburg
models},
JHEP {\bf 0302} (2003) 006;
{\sf arXiv:hep-th/0209098}.

\bibitem{Koerber:2003ef}
P.~Koerber, S.~Nevens and A.~Sevrin,
{\it Supersymmetric non-linear sigma-models with boundaries revisited},
JHEP {\bf 0311} (2003) 066;
{\sf arXiv:hep-th/0309229}.

\bibitem{Sevrin:2007yn}
A.~Sevrin, W.~Staessens and A.~Wijns,
{\it The world-sheet description of A and B branes revisited},
JHEP {\bf 0711} (2007) 06; {\sf arXiv:0709.3733 [hep-th]}.

\bibitem{Sevrin:2008tp}
A.~Sevrin, W.~Staessens and A.~Wijns,
{\it An N=2 worldsheet approach to D-branes in bihermitian geometries: I. Chiral
and twisted chiral fields},
JHEP {\bf 0810} (2008) 108; {\sf arXiv:0809.3659 [hep-th]}.

\bibitem{Sevrin:2009na} 
A.~Sevrin, W.~Staessens and A.~Wijns, 
{\it An N=2 worldsheet approach to D-branes in bihermitian geometries: 
II. The general case}, JHEP {\bf 0909} (2009) 105; 
{\sf arXiv:0908.2756 [hep-th]}.

\bibitem{Dorrzapf:1997rx} 
M.~D\"orrzapf, 
{\it The definition of Neveu-Schwarz superconformal fields and uncharged superconformal transformations}, 
Rev.\ Math.\ Phys.\  {\bf 11} (1999) 137;  
{\sf arXiv:hep-th/ 9712107}.

\bibitem{DeWittBook} B.~DeWitt, {\it Supermanifolds}, 2nd Edition, 
Cambridge University Press (1992).

\bibitem{VoronovBook} T.~Voronov, 
{\it Geometric integration theory on supermanifolds}, 
Sov.\ Sci.\ Rev.\ C.\ Math.\ Phys.\  {\bf 9} (1992) 1.

\bibitem{RogersBook} A.~Rogers, 
{\it Supermanifolds: theory and applications}, 
World Scientific (2007).

\bibitem{Berezin1} 
F.A.~Berezin, 
{\it The method of second quantisation}, Academic, New York (1966).

\bibitem{BerezinLeites}
F.A.~Berezin and D.A.~Le$\hat{\text{i}}$tes,  
{\it Supermanifolds},
Sov.\ Math.\ Dokl.\ {\bf 16} (1975) 1218.

\bibitem{BernsteinLeites1} 
I.N.~Bernstein and D.A.~Le$\hat{\text{i}}$tes, 
{\it Integral forms and Stokes' formula on supermanifolds},
Func.\ Anal.\ Appl.\  {\bf 11} (1977) 45.

\bibitem{BernsteinLeites2} 
I.N.~Bernstein and D.A.~Le$\hat{\text{i}}$tes,  
{\it How to integrate differential forms on supermanifolds},
Func.\ Anal.\ Appl.\  {\bf 11} (1977)  219.

\bibitem{Berezin2} 
F.A.~Berezin, 
{\it Differential forms on supermanifolds}, 
Sov.\ J.\ Nucl.\ Phys.\ {\bf 30} (1979) 605.

\bibitem{Rogers1} 
A.~Rogers, {\it Consistent superspace integration}, 
J.\ Math.\ Phys.\  {\bf 26} (1985) 3. 

\bibitem{Rothstein} 
M.~Rothstein,
{\it Integration on noncompact supermanifolds}, 
Trans.\ Americ.\ Maths.\ Soc.\ {\bf 299} (1987) 387.

\bibitem{Recknagel:2000ri}
A.~Recknagel, D.~Roggenkamp and V.~Schomerus,
{\it On relevant boundary perturbations of unitary minimal models},
Nucl.\ Phys.\  B {\bf 588} (2000) 552; 
{\sf arXiv:hep-th/0003110}.

\bibitem{Recknagel:1997sb}
A.~Recknagel and V.~Schomerus,
{\it D-branes in Gepner models},
Nucl.\ Phys.\  B {\bf 531} (1998) 185;
{\sf arXiv:hep-th/9712186}.

\bibitem{Brunner:1999jq} 
I.~Brunner, M.R.~Douglas, A.E.~Lawrence and C.~R\"omelsberger,
 {\it D-branes on the quintic}, 
 JHEP {\bf 0008} (2000) 015;
 {\sf arXiv:hep-th/9906200}.


\bibitem{Brunner:2005fv}
I.~Brunner and M.R.~Gaberdiel,
{\it Matrix factorisations and permutation branes},
JHEP {\bf 0507} (2005) 012; 
{\sf arXiv:hep-th/0503207}.


\end{thebibliography}
\end{document}